\documentclass[aps,prb,preprint,groupedaddress]{revtex4-1}
\usepackage{graphics}
\begin{document}

% Use the \preprint command to place your local institutional report
% number in the upper righthand corner of the title page in preprint mode.
% Multiple \preprint commands are allowed.
% Use the 'preprintnumbers' class option to override journal defaults
% to display numbers if necessary
%\preprint{}

%Title of paper
%%%%%%%%%%%%%%\title{}
\title{ Kosterlitz-Thouless Phase Transition of the ANNNI model
in Two Dimensions} 
% repeat the \author .. \affiliation  etc. as needed
% \email, \thanks, \homepage, \altaffiliation all apply to the current
% author. Explanatory text should go in the []'s, actual e-mail
% address or url should go in the {}'s for \email and \homepage.
% Please use the appropriate macro foreach each type of information

% \affiliation command applies to all authors since the last
% \affiliation command. The \affiliation command should follow the
% other information
% \affiliation can be followed by \email, \homepage, \thanks as well.
\author{T. Shirakura}%
\email[]{shira@iwate-u.ac.jp}%
\affiliation{%
Faculty of Humanities and Social Sciences,%
Iwate University, Morioka 020-8550, Japan%
}%
\author{F. Matsubara}%
\affiliation{%
Department of Applied Physics, Tohoku University, Sendai 980-8579, Japan%
}%
\author{N. Suzuki}%
%%%%%\email[]{nobu@ait.tbgu.ac.jp}%
\affiliation{%
Faculty of Science and Technology, Tohoku Bunka Gakuen University, Sendai 980-8551, Japan%
}%

%%%%%%%%%%%%%%%%%\author{}
%\email[]{Your e-mail address}
%\homepage[]{Your web page}
%\thanks{}
%\altaffiliation{}
%%%%%%%%%%%%%%%%%%%\affiliation{}

%Collaboration name if desired (requires use of superscriptaddress
%option in \documentclass). \noaffiliation is required (may also be
%used with the \author command).
%\collaboration can be followed by \email, \homepage, \thanks as well.
%\collaboration{}
%\noaffiliation

\date{\today}

\begin{abstract}

The spin structure of an axial next-nearest-neighbor Ising (ANNNI) model in 
two dimensions (2D) is a renewed problem because different Monte Carlo (MC) simulation 
methods predicted different spin orderings.
The usual equilibrium simulation predicts the occurrence of a floating 
incommensurate (IC) Kosterlitz-Thouless (KT) type phase, which never emerges in 
non-equilibrium relaxation (NER) simulations. 
In this paper, we first examine previously published results of both methods, 
and then investigate a higher transition temperature, $T_{c1}$, between the IC and 
 paramagnetic phases. 
In the usual equilibrium simulation, we calculate the chain magnetization   
 on larger lattices (up to $512 \times 512$ sites) and estimate 
$T_{c1} \approx 1.16J$ with frustration ratio $\kappa (\equiv -J_2/J_1) = 0.6$. 
We examine the nature of the phase transition in terms of the Binder ratio $g_L$ 
of spin overlap functions and the correlation-length ratio $\xi/L$. 
In the NER simulation, we observe the spin dynamics in equilibrium 
states by means of an autocorrelation function, and also observe  
the chain magnetization relaxations from the ground and disordered states. 
These quantities exhibit an algebraic decay at $T \lesssim 1.17J$. 
We conclude that the two-dimensional ANNNI model actually admits
an IC phase transition of the KT type. 

\end{abstract}


\maketitle

%%%%%%%%%%%%%%%%%%%%%%%%%%%%%%%%%%%%%%%%%%%%%%%%%%%%%%%%%%%%%%%%%%%%%%%%%%%%%%%%%

\section{Introduction}

Systems with competitive interactions have been extensively studied throughout
 the past three decades, because they exhibit rich physical phenomena, 
 such as commensurate-incommensurate phase transitions, 
Lifshitz points, and multiphase points.\cite{Selke0} 
The axial next-nearest-neighbor Ising (ANNNI) model is among the 
simplest realizations of such systems. 
In the two-dimensional (2D) ANNNI model, ferromagnetic Ising chains are coupled 
by ferromagnetic nearest-neighbor and antiferromagnetic 
next-nearest-neighbor interchain interactions on a square lattice. 
The Hamiltonian is described by
\begin{eqnarray}
\mathcal{H} &=& - \it{J}\sum_{\langle\it{x,y}\rangle}S_{\it{x,y}}S_{\it{x+1,y}} \nonumber \\
     &-& \it{J_1}\sum_{\langle\it{x,y}\rangle}S_{x,y}S_{x,y+1}
     - \it{J}_{\it{2}}\sum_{\langle\it{x,y}\rangle}S_{x,y}S_{x,y+2},
\end{eqnarray}
where $S_{x,y}=\pm 1$ is an Ising spin. In this paper we consider the case 
with $J_1\; =\; J\; >0$ and $J_2 < 0$. 
The ground state of the model is a ferromagnetic phase for frustration coefficient
$\kappa (\equiv -J_2/J) <1/2$ 
and an antiphase ($\langle 2 \rangle$ phase)  for $\kappa > 1/2$. 
This state is described by an alternate arrangement of two up-spin and two down-spin 
chains in the $y$-direction. 
This model at finite temperatures has been studied throughout the past few decades. 
At high temperatures and $\kappa < 1/2$, the model transits 
from the ferromagnetic phase to a paramagnetic (PM) phase. 
On the other hand, the spin structure for $\kappa > 1/2$ is yet to be clarified. 
Early Monte Carlo (MC) simulations suggested that a floating 
incommensurate (IC) phase exists between the $\langle 2 \rangle$ phase and the PM 
phase.\cite{Selke1,Selke2}
Furthermore, the IC phase close to the higher transition temperature, $T_{c1}$,  
may be characterized by dislocations that
play the same role of vortices in two-dimensional XY (2D XY) model.\cite{Selke1,Selke2} 
Since the phase transition at $T_{c1}$ is considered equivalent to the 
Kosterlitz-Thouless\cite{KT} (KT) type in the 2D XY model, it is called 
the KT phase transition. This picture of the spin ordering has been supported by various
theoretical\cite{Villain,Grynberg1} and approximation\cite{Saqi,Morita} studies. 
Sato and Matsubara (SM)\cite{Sato} simulated an equilibrium scenario using a cluster heat 
bath (CHB) algorithm.\cite{CHB1,CHB2} They found that as the temperature is lowered, 
the KT transition yields the IC phase at $T_{c1}$ and 
 the $\langle 2\rangle$ phase at temperature $T_{c2}$. The estimated transition temperatures
were $T_{c1} \approx 1.16J$ and $T_{c2} \approx 0.91J$  at $\kappa = 0.6$. 
On the other hand, Shirahata and Nakamura (SN)\cite{Shirahata} investigated the spin 
ordering of the same model using a nonequilibrium relaxation (NER) method\cite{ItoA,ItoB} 
and reported $T_{c1} \approx 0.89J$ and $T_{c2} \approx 0.89J$ for $\kappa = 0.6$.
Rastelli et. al.\cite{Rastelli} conducted the equilibrium MC simulation using 
a single-spin-flip algorithm with a huge number of MC sweeps ($10^7 - 10^8$) 
and obtained $T_{c1} \approx 1.27J$ and $T_{c2} \approx 0.95J$. The NER simulations of
Chandra and Dasgupta\cite{Chandra} yielded 
$T_{c1} = T_{c2} \approx 1.00J$.
Clearly, the presence of the IC phase depends on the simulation method; the IC phase
emerges in equilibrium simulations but 
is absent in NER simulations. 

To confirm the conclusions of these simulation methods,
we must question their implementation. 
The equilibrated system
in the equilibrium simulation is moderately small, 
occupying up to $64 \times 128$ sites\cite{Sato} or $96 \times 96$ sites.\cite{Rastelli} 
Is this system size sufficiently large to predict the phase transition of the model? 
Although the system size is much larger in NER simulations,  
(typically $1999 \times 2000$ sites), the initial stage 
 of the MC simulation is limited to approximately $10^5$ MC sweeps.
In complex systems 
with very slow relaxation, 
is this initial relaxation phase sufficiently slow to capture the critical relaxation? 

In this paper we reexamine the existence of the IC phase in the ANNNI model with 
$\kappa = 0.6$ by conducting both equilibrium and NER simulations.
Since both simulation methods predict the phase transition at $T_{c2}$
 ( approximately  $0.9J)$,\cite{Sato,Shirahata} 
 we focus on the occurrence of the IC phase transition at 
$T_{c1} ( > T_{c2})$. 
In the equilibrium simulation, we extend the lattice size up to 
$512 \times 512$ sites to examine the size effect. In the NER simulation, we examine 
the equilibrium process during a long MC run. 
We also calculate the autocorrelation function of the equilibrium state in the IC phase. 
Besides the chain magnetization in the $x$-direction, we consider the spin overlap of 
two replicas which is usually investigated in the spin glass problem.
The investigated methods and physical quantities are described
in Section II.
Section III presents the results of the equilibrium simulation.
In Section IV, first we examine the results of recent NER simulations, 
and then we investigate 
the equilibrium process of the model assuming as initial configurations 
in both the $\langle 2 \rangle$ phase and the PM phase. 
We also investigate the dynamical property of this model in 
the equilibrium state. Conclusions are presented in Section V.

%%%%%%%%%%%%%%%%%%%%%%%%%%%%%%%%%%%%%%%%%%%%%%%%%%%%%%%%%%%%%%%%%%%%%%%%

\section{Methods and Quantities}

An ANNNI model with $\kappa = 0.6$ was set up on $L_0 \times L_0$ lattices 
with open boundary conditions in both $x$- and $y$-directions. 
These boundary conditions naturally reflect the surfaces of real materials.
 Open boundaries release the relaxation time in slow relaxation systems.\cite{Sato}  
 We measured the physical quantities of interest 
in the inner regions, which are not subject to surface effects.
 The linear size $L$ of the measuring region was varied with 
$L_n (\equiv L_0/2^n)$ (n=0,1 and 2) (see Fig. 1). 
Two MC algorithms were applied in our simulation.
%%%%%%%%%%%%%%%%%%%%%%%%%%%%%%%%%%%%%%%%%%%%%%%%%%%%%
\begin{figure}[tbhp]
\begin{center}
\includegraphics{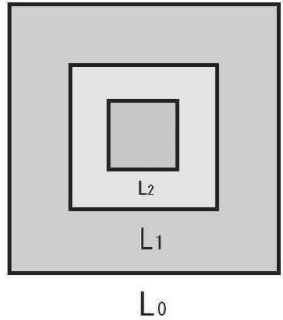}\\
\end{center}
\vspace{-0.4cm}
\caption{
An open-boundary square lattice of linear size $L_0$ and its nested inner regions 
(linear sizes $L_1 = L_0/2$ and $L_2 = L_0/4$).
}
\end{figure}
%%%%%%%%%%%%%%%%%%%%%%%%%%%%%%%%%%%%%%%%%%%%%%%%%%%%%%

\noindent
A) Single-Spin-Flip (SSF) algorithm 

Because the NER method is based on the SSF dynamics, we study the NER of
order parameters using 
 a conventional SSF heat bath algorithm.
 
\noindent
B) The CHB algorithm

We use the CHB algorithm in the equilibrium simulation because 
this algorithm reduces the number of MC sweeps in the relaxation.  In the CHB algorithm, 
the spin configuration of a block of $L_0 \times l_y$ spins is updated 
using the transfer matrix method,
where the transfer direction is the $x$-direction ($L_0$) and $l_y$ is determined 
from the computational time costs. In this paper we apply the SM procedure\cite{Sato} 
with $l_y=6$.

We consider two quantities: the square of the chain magnetization (the magnetization 
along the $x$-axis) given by 
\begin{eqnarray}
	M_2 = \frac{1}{L} \sum _{y=1}^{L} (\frac{1}{L} \sum _{x=1}^{L} S_{x,y})^2. 
\end{eqnarray}
$M_2$ is conventionally used to examine the phase transition of the model. 
We also consider the spin overlap function $q(\vec{k})$ of two spin 
configurations $\{S_{x,y}^{\alpha}\}$ and $\{S_{x,y}^{\beta}\}$ in independent MC runs:  
\begin{eqnarray}
q(\vec{k})=\frac{1}{L^2}\sum_{x=1}^{L}\sum_{y=1}^{L}S_{x,y}^{\alpha}S_{x,y}^{\beta}
\exp(i\vec{k}\vec{r}_{x,y}).
\end{eqnarray}
From the $q(\vec{k})$, we investigate the nature of the phase transition.

%%%%%%%%%%%%%%%%%%%%%%%%%%%%%%%%%%%%%%%%%%%%%%%%%%%%%%%%%%%%%%%%%%%%%%%%%%%%%%%%
%%%%%%%%%%%%%%%%%%%%%%%%%%%%%%%%%%%%%%%%%%%%%%%%%%%%%%%%%%%%%%%%%%%%%%%%%%%%%%%%

\section{Equilibrium Simulation} 

We investigate the equilibrium properties of the ANNNI model by 
 SM's approach.\cite{Sato} Especially, we are interested in
 whether the previous picture 
of the spin ordering emerges on larger lattices. 
Therefore, we extend the lattice size to the largest possible, 
up to $L_0 \times L_0 = 512 \times 512$ sites, with linear size eight 
times larger than that treated by SM.  

The physical quantities were averaged from 16 independent simulation runs. 
The system is regarded as equilibrated when the difference 
in $\langle q(0)^2\rangle$ between two estimates obtained from two 
different MC sweeps 
(the one averaged over from $MCS_{equi}+1$ MC sweep to $MCS_{equi}+MCS_{mea}/2$
MC sweep and the other from $MCS_{equi}+MCS_{mea}/2+1$ MC sweep 
to $MCS_{equi}+MCS_{mea}$ MC sweep) 
becomes smaller than $1\%$, 
where $\langle ...\rangle$  denotes the thermal average. 
About 80,000 MC sweeps were needed to equilibrate the system with $L_0=512$ at
 $T=1.15J$. The parameters used in the equilibrium simulation are listed in 
Table~\ref{tab:table1}.

%%%%%%%%%%%%%%%%%%%%%%%%%%%%%%%%%%%%%%%%%%%%%%%%%%%%%%
\begin{table}[b]
\caption{\label{tab:table1}
Parameters used in the CHB algorithm of the MC simulation. $MCS_{equi}$ and $MCS_{mea}$ 
are the number of MC sweeps required for equilibration and measurement, respectively.}
\begin{ruledtabular}
\begin{tabular}{rrr}
$L_0$ & $MCS_{equi}$ & $MCS_{mea}$ \\ 
\hline
  32 &  4,000 & 12,000 \\ 
  64 &  10,000 & 30,000 \\
 128 &  20,000 & 60,000 \\ 
 256 &  40,000 & 80,000 \\
 512 &  80,000 & 120,000 \\
\end{tabular}
\end{ruledtabular}
\end{table}
%%%%%%%%%%%%%%%%%%%%%%%%%%%%%%%%%%%%%%%%%%%%%%%%%%%%%%%%%%%%%%%%%%%%%

\subsection{Chain magnetization}

%%%%%%%%%%%%%%%%%%%%%%%%%%%%%%%%%%%%%%%%%%%%%%%
\begin{figure}
\includegraphics{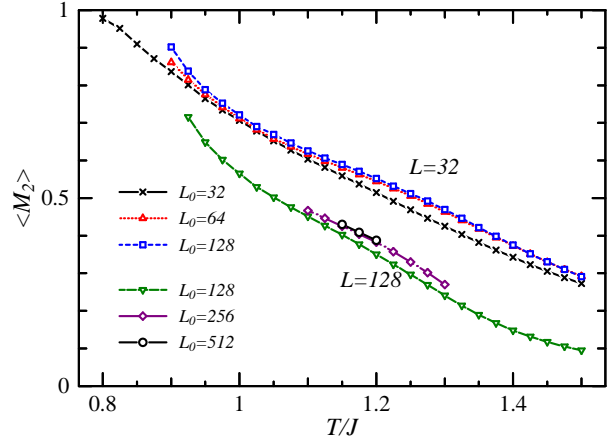}%
\vspace{-0.4cm}
\caption{(color online) Lattice size $L_0$ dependences of $\langle M_2\rangle$ 
in the ANNNI model with $\kappa = 0.6$ within regions $L = 32$ and $L = 128$. 
Error bars are smaller than the symbols.
}
\end{figure}
%%%%%%%%%%%%%%%%%%%%%%%%%%%%%%%%%%%%%%%%%%%%%%%%%%%%%%

%%%%%%%%%%%%%%%%%%%%%%%%%%%%%%%%%%%%%%%%%%%%%%%%%%%%%
\begin{figure}
\includegraphics{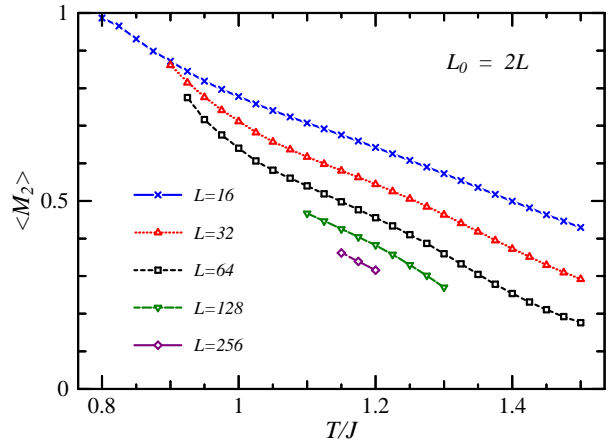}
\vspace{-0.4cm}
\caption{(color online) 
Temperature dependences of the square of the chain magnetization $\langle M_2\rangle$ 
in the ANNNI model with $\kappa = 0.6$ within regions of different linear size $L$.
Error bars are smaller than the symbols.
}
\end{figure}
%%%%%%%%%%%%%%%%%%%%%%%%%%%%%%%%%%%%%%%%%%%%%%%%%%%%%%%

We first consider the square of the chain magnetization $\langle M_2\rangle$. 
To examine the surface effect,
 we plot $\langle M_2\rangle$ as a function of temperature within the inner regions
$L = 32$ and  $L=128$ at different lattice sizes $L_0$.
The results are shown in Fig. 2. We find that 
 $\langle M_2\rangle$ increases with increasing $L_0$. 
Although $\langle M_2\rangle$ markedly differs between the whole lattice 
 $L_0$ and the inner region $L_1$, it little varies between 
the inner regions $L_1$ and $L_2$. 
Therefore, we consider that the surface effect can be eliminated by conducting 
simulations over the inner region $L_1$. 
Figure 3 plots $\langle M_2\rangle$ in the inner region $L_1$  as a function of
temperature. We investigate 
the phase transition in the model identically to SM.\cite{Sato} 
If the IC phase occurs, the spin correlation function will decay according to a power law 
%%%%%%%%%%%%%%%%%%%%%%%%%%%%%%%%%%%%%%%%%%
\begin{eqnarray} 
\langle S_{\it{0,0}} S_{\it{x,y}}\rangle \sim r^{-\eta}\cos(qy) \hspace{0.5cm} {\rm for} 
\hspace{0.5cm} x, y \gg 1
\end{eqnarray} 
%%%%%%%%%%%%%%%%%%%%%%%%%%%%%%%%%%%%%%%%%%
where $r = \sqrt{x^2+y^2}$ and $\eta$ and $q$ are the decay exponent and the wave number, 
respectively. 
At the transition temperature $T_{c1}$ or below,  
the chain magnetization is described by
%%%%%%%%%%%%%%%%%%%%%%%%%%%%%%%%%%%%%%%%%
\begin{eqnarray}
\langle M_2\rangle 
&=& \frac{1}{L}\sum_{y=1}^{L}\frac{1}{L} \sum_{x=1}^{L} \langle S_{1,y}S_{x,y} \rangle \\
&\sim& L^{-\eta}. 
\end{eqnarray}
%%%%%%%%%%%%%%%%%%%%%%%%%%%%%%%%%%%%%%%%%
 First, we examine this relationship. 
By tuning $\eta$, all of the $\langle M_2\rangle L^\eta - T/J$ curves can be made to cross 
at a single point. From this crossover point, $T_{c1}$ and $\eta$ are determined as 
approximately $1.16J$ and 0.25, respectively. 
The result is plotted in Fig. 4. 
Next we construct a finite-size scaling (FSS) plot, assuming the KT transition\cite{KT} 
%%%%%%%%%%%%%%%%%%%%%%%%%%%%%%%%%%%%%%%%%%
\begin{eqnarray}
\langle M_2\rangle L^\eta = Y[L^{-1}\exp(b|\varepsilon|^{-0.5})],
\end{eqnarray}
%%%%%%%%%%%%%%%%%%%%%%%%%%%%%%%%%%%%%%%%%%
where $\varepsilon=(T-T_{c1})/T_{c1}$ and $Y$ is some scaling function. 
Setting $T_{c1}$ and $\eta$ to approximately $1.16J$ and 0.25, respectively,
 and $b=2.2$, 
the curves neatly collapse at the higher temperature side $T > T_{c1}(= 1.16J)$, 
as shown in Fig. 5. 
However, the FSS plots fail at the lower temperature side $T < T_{c1}$, 
 implying that no long range order occurs at $T < T_{c1}$. 
%%%%%%%%%%%%%%%%%%%%%%This problem will be discussed in the next subsection. 
We should note that the values estimated here are consistent with 
those estimated by SM on smaller lattices ($L_0 \le 64$), 
reported as $T_{c1}/J = 1.16 \pm 0.02, \eta = 0.25 \pm 0.02$ and $b \sim 2.2$.\cite{Sato} 
  
%%%%%%%%%%%%%%%%%%%%%%%%%%%%%%%%%%%%%%%%%%%%%%%%%%%%%
\begin{figure}
\includegraphics{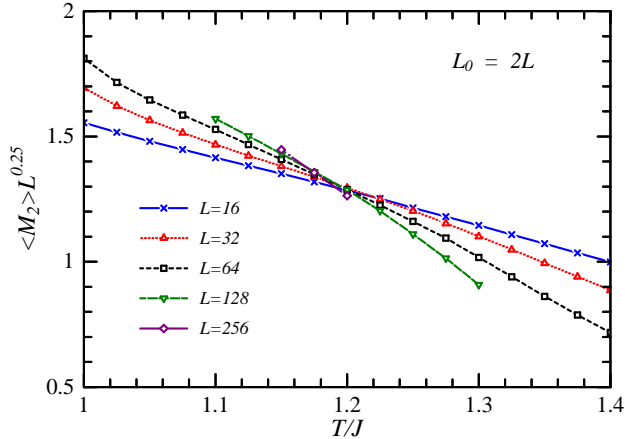}\\
\vspace{-0.4cm}
\caption{(color online) 
$\langle M_2\rangle L^{0.25}$ versus $T/J$.
}
\end{figure}
%%%%%%%%%%%%%%%%%%%%%%%%%%%%%%%%%%%%%%%%%%%%%%%%%%%%%%
  
%%%%%%%%%%%%%%%%%%%%%%%%%%%%%%%%%%%%%%%%%%%%%%%%%%%%%
\begin{figure}
\includegraphics{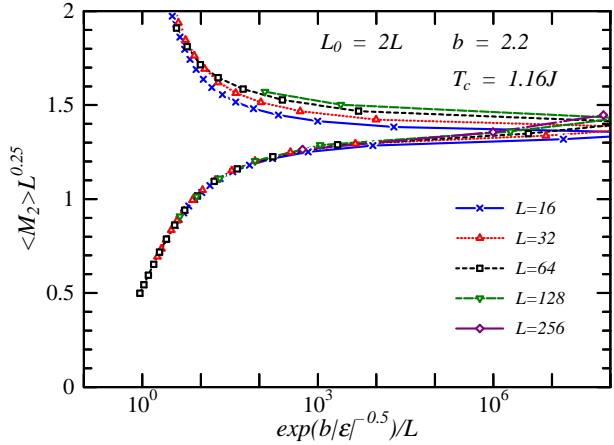}\\
\caption{ (color online) 
Finite-size scaling plots of the ANNNI model with $\kappa = 0.6$, 
assuming the KT transition with the same parameters as SM.\cite{Sato}
}
\end{figure}
%%%%%%%%%%%%%%%%%%%%%%%%%%%%%%%%%%%%%%%%%%%%%%%%%%%%%%

\subsection{Spin overlap}

We now consider the spin overlap function. 
 The temperature dependence of $\langle q(0)^2 \rangle$ over the whole 
lattice $L_0$ is plotted in Fig. 6. 
We find that $\langle q(0)^2 \rangle$ is a decreasing function of temperature. 
Efficient methods have been developed 
for determining the transition temperature from the spin overlap function. 
Here, we apply these methods to investigate the phase transition. 
However, these methods examine the ratios of the moments of 
the spin overlap functions 
which yield scattered data.
We then consider the spin overlap functions in the inner region with $L = L_2$.

%%%%%%%%%%%%%%%%%%%%%%%%%%%%%%%%%%%%%%%%%%%%%%%%%%%%%
\begin{figure}
\includegraphics{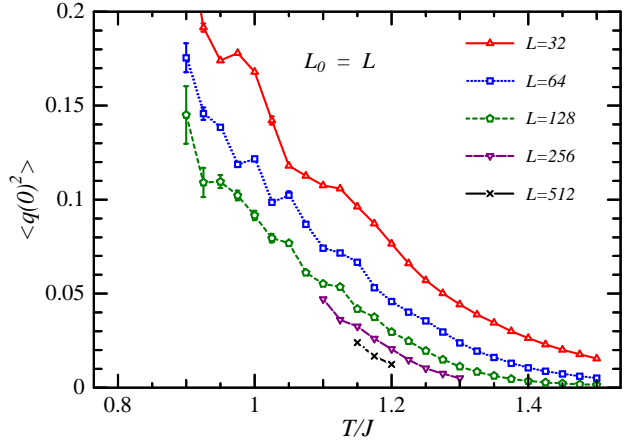}\\
\vspace{-0.4cm}
\caption{ (color online) 
Temperature dependences of the spin overlap function $\langle q(0)^2 \rangle$ 
in the ANNNI model with $\kappa = 0.6$, computed over the whole lattice with $L = L_0$.
Error bars that are not shown are smaller than the symbols.
}
\end{figure}
%%%%%%%%%%%%%%%%%%%%%%%%%%%%%%%%%%%%%%%%%%%%%%%%%%%%%%

%%%%%%%%%%%%%%%%%%%%%%%%%%%%%%%%%%%%%%%%%%%%%%%%%%%%%
\begin{figure}
\includegraphics{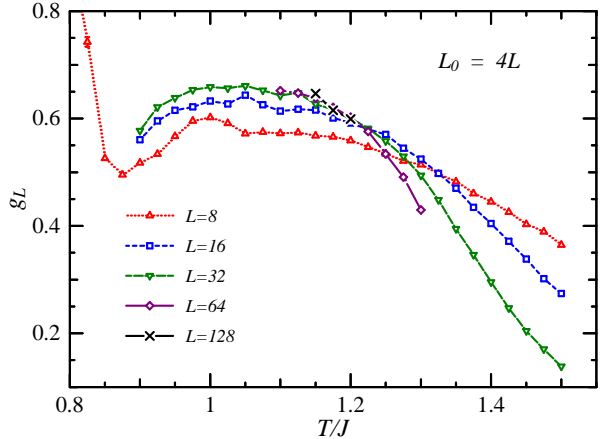}\\
\caption{(color online) 
Temperature dependences of Binder ratio $g_L$ in the ANNNI model with $\kappa = 0.6$ 
and different $L$. Error bars are smaller than the symbols.
}
\end{figure}
%%%%%%%%%%%%%%%%%%%%%%%%%%%%%%%%%%%%%%%%%%%%%%%%%%%%%%

First we consider the Binder ratio $g_L$\cite{Binder} defined as
%%%%
\begin{eqnarray}
g_L = \frac{1}{2}\left(3-\frac{\langle q(0)^4\rangle}{\langle q(0)^2\rangle^2}\right).  
\end{eqnarray}
%%%%
The temperature dependences of $g_L$ for different $L$ are plotted in Fig. 7. 
At high temperatures, $g_L$ decreases with increasing $L$, indicating that no long range order establishes at these temperatures. 
As the temperature is decreased, $g_L$ for larger $L$ 
%%increases by increments that depend on $L$ (the higher the $L$, 
%%the larger the increment).  $g_L$ and $g_{2L}$ converge
%% at $T(L)$. The temperature $T(L)$ decreases as $L$ increases and appears 
%% to converge at approximately $T \approx 1.18J$,  close to the transition 
%%temperature $T_{c1}$ estimated in Sec. III A.
converge at $T \approx 1.18J$. 
%%%,  close to the transition temperature $T_{c1}$ estimated in Sec. III A.
Therefore, the Binder ratio $g_L$ supports that a phase transition 
occurs at $T = T_{c1}$. 
Below this temperature, the $L$ dependence of $g_L$ differs from that 
of usual systems exhibiting long-range order at low temperatures. 
That is, $g_L$ slightly increases with increasing $L$ and appears to converge 
to a single line. 
An analogous phenomenon occurs in the 2D XY model,\cite{2DXYBind} 
indicating that the IC phase at $T \lesssim T_{c1}$ is indeed a KT type phase. 
Another remarkable feature is the behavior of $g_L$ as the temperature falls 
below $T_{c1}$; $g_L$ slightly increases, is maximized at 
$T \approx 1.05J$, and decreases below $T \approx 1.00J$. 
This behavior may imply that a different spin correlation develops below 
$T \approx 1.05J$. 
As is well-known, slightly above the lower transition temperature $T_{c2}$ 
the spin structure of the IC state is characterized by domain walls of three 
up-spin or down-spin chains that penetrate the 
$\langle 2 \rangle$ phase.\cite{Selke2}
%%%Therefore, $T \approx 1.05J$ is a crossover temperature around which 
%%%the IC state of the KT type at $T \lesssim T_{c1}$ gradually changes 
%%%to the low temperature IC state characterized by the domain walls. 

%%%%%%%%%%%%%%%%%%%%%%%%%%%%%%%%%%%%%%%%%%%%%%%%%%%%%
\begin{figure}
\includegraphics{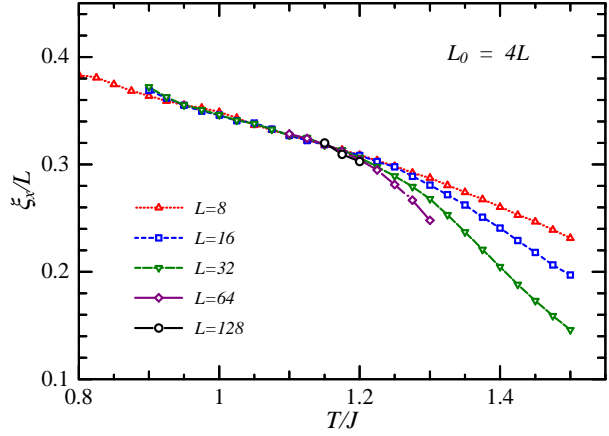}\\
\caption{ (color online) 
Correlation length ratio $\xi_x/L$ in the ANNNI model with $\kappa = 0.6$ 
in the x direction. Error bars are smaller than the symbols.
}
\end{figure}
%%%%%%%%%%%%%%%%%%%%%%%%%%%%%%%%%%%%%%%%%%%%%%%%%%%%%%%%%%%%%%%%%%%%

%%%%%%%%%%%%%%%%%%%%%%%%%%%%%%%%%%%%%%%%%%%%%%%%%%%%%
\begin{figure}
\includegraphics{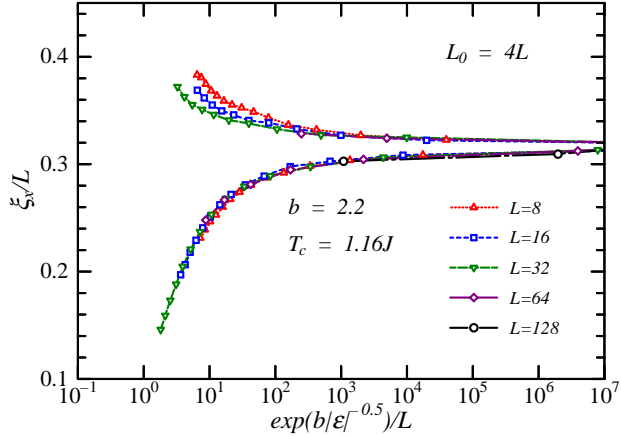}\\
\vspace{-0.4cm}
\caption{ (color online) 
Finite-size-scaling plot of $\xi_x/L$ in the region $L = L_2$ assuming 
$T_{c1} = 1.16J$. 
}
\end{figure}
%%%%%%%%%%%%%%%%%%%%%%%%%%%%%%%%%%%%%%%%%%%%%%%%%%%%%%%%%%%%%%%%%%%%

\subsection{Correlation length}

Next we consider the spin correlation length $\xi_{\mu}$ ($\mu = x, y$) along 
the $\mu$-direction. This quantity is obtained from the spin overlap function as follows:  
\begin{eqnarray}
\xi_{\mu} = \frac{1}{2\sin(|\vec k_{\rm min}|/2)}
        \sqrt{\frac{\langle q(0)^2\rangle}{\langle |q({\vec k}_{\rm min})|^2\rangle}-1}
\end{eqnarray}
where $ {\vec k}_{\rm min} = (\pi/L, 0)$ and $ {\vec k}_{\rm min} = (0, \pi/L)$ in the $x$- and $y$-direction, respectively. 
The ratio of the correlation length $\xi_{\mu}$ to the linear lattice size 
$L$,  $\xi_{\mu}/L$, determines the transition temperature 
$T_c$.\cite{Cooper}  When $T > T_c$, $\xi_{\mu}$ is finite and $\xi_{\mu}/L 
\rightarrow 0$ as $L \rightarrow \infty$. 
On the other hand, at $T = T_c$, $\xi_{\mu}$  diverges in the thermodynamic 
limit and $\xi_{\mu}/L = C (\neq 0)$.
Therefore, the $\xi_{\mu}/L$ for different $L$ cross at the phase transition 
temperature $T_c$. 
The correlation-length ratios $\xi_x/L$ for different $L$ 
are plotted as functions of $T$ in Fig. 8. 
At high temperatures, $\xi_x/L$ reduces at larger $L$, indicating that 
no long-range order establishes at these temperatures. 
As the temperature is decreased, the $\xi_x/L$ values 
 increase for all $L$, and converge at approximately $T \approx 1.17J$. 
Below this temperature, they slowly increase at the same rate. 
This behavior is also observed in the 2D XY model.\cite{2DXYCorr} 
To estimate the transition temperature $T_{c1}$, we construct an FSS plot of 
the $\xi_x/L$ values. The FSS plot collapses above 
$T_{c1}$, when $T_{c1}$ is assumed as $1.16J$(see Fig. 9).

%%%%%%%%%%%%%%%%%%%%%%%%%%%%%%%%%%%%%%%%%%%%%%%%%%%%%
\begin{figure}
\includegraphics{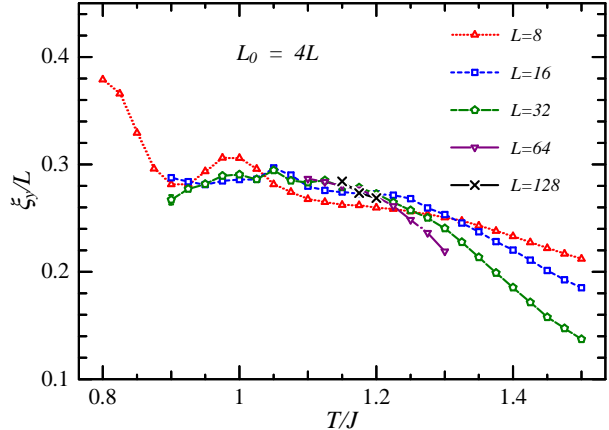}\\
\vspace{-0.4cm}
\caption{ (color online) 
Correlation length ratio $\xi_y/L$ in the ANNNI model with $\kappa = 0.6$ 
in the y direction. Error bars are smaller than the symbols.
}
\end{figure}
%%%%%%%%%%%%%%%%%%%%%%%%%%%%%%%%%%%%%%%%%%%%%%%%%%%%%%

Figure 10 plots the correlation-length ratio $\xi_y/L$ along the $y$-axis. 
Identically to their  $\xi_x/L$ counterparts, 
the $\xi_y/L$ values for different $L$ converges at $T \approx 1.17J$. 
%Below this temperature, $\xi_y/L$ becomes a function 
%of temperature, although 
%%%% alone(rather than dependent both $L$ and $T$). 
%%%Two features of these graphs should be noted. First, 
%the data are scattered for small $L$. 
%%%% because of  competition by interactions 
%%%$J( > 0)$ and $J_2 ( < 0)$ in the $y$-direction. 
As the temperature decreases below $T_{c1}$, $\xi_y$ first slightly 
increases down to $T \approx 1.05J$, and slightly decreases thereafter,
except for the data of $L=8$. 
This temperature dependence of $\xi_y$ at $T < 1.05J$ differs from that of 
$\xi_x$. Specially, at $T < 1.05J$, the spin correlations 
in the $x$-direction are enhanced as 
the temperature decreases, while those in the $y$-direction are suppressed. 
%This result provides evidence of the development of the low temperature 
%spin correlation. 

\subsection{Summary}
 
We have investigated the phase transition in the 2D ANNNI model by conducting 
equilibrium MC simulations. 
We calculated the square of the chain magnetization in larger lattices of 
$L_0 \times L_0$ sites ($L_0 \le 512$) and obtained $T_{c1}\approx 1.16J$, 
absolutely consistent with the results of the previous simulations on small 
lattices ($L_0 \le 64$). 
Thus, we conclude that the IC phase actually occurs in the ANNNI model.  

We also calculated the Binder ratio $g_L$ of the spin overlap functions 
 and the correlation-length ratios $\xi_x/L$ and $\xi_y/L$. 
At $T \lesssim T_{c1}$, these quantities behave similar to 
those in the 2D XY model. This suggests an analogy between
 the IC phase in the ANNNI model at $T \lesssim T_{c1}$ and 
 the Kosteritz Thouless (KT) phase in the 2D XY model. 
Therefore, we can naturally refer to the phase transition at $T = T_{c1}$ as
the KT phase transition.

%%%%%%%%%%%%%%%%%%%%%%%%%%%%%%%%%%%%%%%%%%%%%%%%%%%%%%%%%%%%%%%%%%%%%%%%%%

\section{NER Simulation}

We now examine previous results of NER simulations.
The NER method is based on the following hypothesis.\cite{ItoA,ItoB} 
In a system with a relevant order parameter $Q$ and a perfectly ordered
initial state $Q(0) = 1$ (or the PM phase $Q(0) = 0$),
 MC simulations on 
a large lattice at temperature $T$ lead to three behaviors in the limit 
 $t \rightarrow \infty$; 
(i) if $T > T_c$, $Q(t)$  decays exponentially, 
(ii) if $T < T_c$, $Q(t)$ converges toward some non-zero value, and 
(iii) if $T = T_c$, $Q(t)$ exhibits an algebraic decay (or an algebraic growth).
In a critical state such as the KT phase,  
$Q(t)$ exhibits a behavior similar to that at $T = T_c$.

Shirahata and Nakamura(SN)\cite{Shirahata} used 
the square of the chain magnetization $m_l(t)\ (\ \equiv \ M_2 $ 
at $t$ MC sweep) as an order parameter of the IC phase. 
They performed MC simulations of the model with $\kappa = 0.6$ starting with both 
the $\langle 2 \rangle$ phase of $m_l(0) = 1$ and the PM phase of $m_l(0) = 0$. 
In the former case,
they found that $m_l(t)$ decays exponentially at $T > 0.98J$; in the latter,
it tends to saturate  
at $T > 0.92J$. From these results they predicted that $T_{c1} < 0.92J$. 
Applying a finite time scaling analysis they refined this result to
 $T_{c1} \sim 0.89J$, close to the $\langle 2 \rangle$ phase 
transition temperature $T_{c2} \sim 0.89J$ estimated from finite time scaling 
analysis of the $\langle 2 \rangle$ phase magnetization. Similarly,
Chandra and Dasgupta(CD)\cite{Chandra} found that  
the order parameter $m_l(t)$ algebraically decays  
at $T \approx 1.00J$. Their transition temperature $T_{c1} \approx 1.00J$ and 
$T_{c2} \approx 1.00J$ (the latter estimated from relaxation of the energy)
are also extremely close. 
Therefore, the NER method predicts the absence of the IC phase.

Besides the considerably different values of 
$T_{c1}$ (approximately $T_{c2}$) between estimated by SN and CD, 
the NER method raises some pertinent issues:
(i) The exponential decay of $m_l(t)$ suggests that only the $\langle 2 \rangle$ phase 
is unstable; it does not reveal the instability of the IC phase. 
In fact, $m_l(t)$ rebounds as the simulation proceeds.\cite{Chandra} Rather,
the stability of the IC phase should be examined by 
 relaxation from an equilibrium state in the IC phase at $T_{c2} < T < T_{c1}$ 
(if present); 
(ii) The initial growth results of $m_l(t)$ reported by SN\cite{Shirahata}
 are not convincing. 
As seen in Fig. 3, the equilibrium value of $m_l(t)$, $\langle M_2\rangle$, 
is higher for the small $L$ than for the large $L$. 
However, $m_l(t)$ depicted in Fig. 6 of SN\cite{Shirahata} is independent of the linear 
lattice size $L_x$ at $t < 10^4$ and increases with $L_x$ at $t > 10^4$.
We consider that the growth of $m_l(t)$ from the PM phase should be reexamined.

Here, we consider two phenomena: (i) The ordering process of the system  
initialized to non-equilibrium states and 
(ii) The dynamics of the system in the equilibrium state. 
%%%The relaxation of the system starting with an equilibrium state. 
Equivalently, we investigate the autocorrelation function in the equilibrium 
state. Since a huge number of MC sweeps are required 
to reach equilibrium, we implement the system on small lattices 
($L_0 \leq 512$).

%%%%%%%%%%%%%%%%%%%%%%%%%%%%%%%%%%%%%%%%%%%%%%%%%%%%%
\begin{figure}
\vspace{0.4cm}
\includegraphics{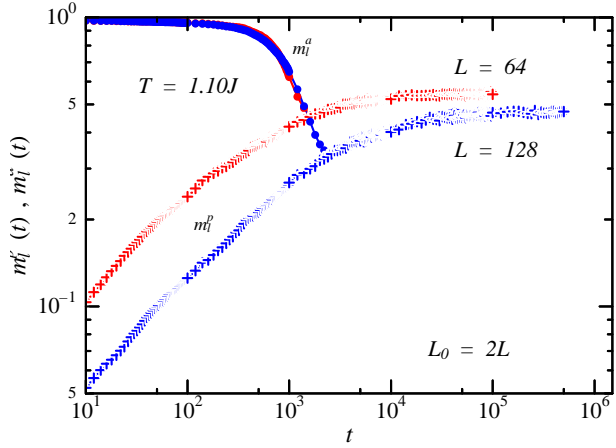}\\
\vspace{-0.4cm}
\caption{ (color online) Relaxations of $m_l^a(t)$ and $m_l^p(t)$ 
in the ANNNI model with $\kappa = 0.6$ at $T = 1.10J$, slightly lower 
than $T_{c1} = 1.16J$ estimated in the equilibrium simulation. }
\end{figure}
%%%%%%%%%%%%%%%%%%%%%%%%%%%%%%%%%%%%%%%%%%%%%%%%%%%%%%%%%%%%%%%%%%%%

\subsection{Relaxations from the $\langle 2 \rangle$ phase and the PM phase}

Starting with the $\langle 2 \rangle$ phase and the PM phase,
we investigate the relaxation of the system. The system is implemented
on the lattice described in Sec. II with 
 $N ( = 64 \sim 256)$ sets of spin configurations. At each MC sweep $t$,
the square of the chain magnetization $m_l^{\mu}(t)$ is computed:  
%%%%%%%%%%
\begin{eqnarray}
	m_l^{\mu}(t) = \overline{M_2(t)},
\end{eqnarray}
%%%%%%%%&
where $M_2(t)$ is defined by eq.(2) at MC sweep $t$ and $\overline{\cdots}$ 
is the configuration average. The PM phase and the $\langle 2 \rangle$ phase
initial states are distinguished by setting the
superscript $\mu = p$ and $a$, respectively.

 Figure 11 plots the time courses of $m_l^p(t)$ and $m_l^a(t)$ calculated by
 the model on lattices with 
$L_0 = 128$ and $L_0 = 256$ at $T = 1.10J$. 
Initially, $m_l^p(t)$ grows while $m_l^a(t)$ decays. 
At later times, the two quantities exhibit quite different temporal behaviors. 
While $m_l^p(t)$ 
monotonically increases and eventually saturates, 
$m_l^a(t)$ rapidly decreases, intercepts $m_l^p(t)$, 
and then increases along it. 
This rebound of $m_l^a(t)$ has been previously reported by CD.\cite{Chandra} 
Importantly, $m_l^a(t)$ saturates at a much higher value 
than its minimum, and the minimum and saturation levels widen with increasing $L$. 
That is, the $\langle 2 \rangle$ phase breaks once and a spin correlation of 
the IC state develops. 
The existence of the IC phase is examined by  
the equilibrium simulation performed in Sec. III. 
Another notable behavior is the large $L$-dependence of $m_l^p(t)$,
which strongly contrasts with the SN results.\cite{Shirahata} 
 This behavior is reasonable because the square 
of the chain magnetization $m_l^{\mu}$ is 
related to the chain susceptibility $\chi_l^{\mu}(t)$  by
%%%%%%%%%%%%%%%%%%%%%%%%%%%%%%%%%%%%%%%%%%%%%%%%%%
\begin{eqnarray}
  m_l^{\mu}(t)  &=& \frac{1}{L} \chi_l^{\mu}(t). 
\end{eqnarray}
%%%%%%%%%%%%%%%%%%%%%%%%%%%%%%%%%%%%%%%%%%%%%%%%%%%
If the spin correlations are not extensively developed, 
 the chain susceptibility should become independent of $L$ at 
large $L$ and thereby reveal the NER properties of the system.
The time courses of the susceptibility $\chi_l^{p}(t) ( \equiv m_l^p(t)L)$
for different $L$ are plotted at $T = 1.10J$ and $T = 1.25J$
in Figs.12 (a) and (b), 
 respectively. Note that 
 as $L$ increases, $\chi_l^p(t)$ converges  at small $t$. 
Following SN, we take $\chi_l^p(t)$ in the thermodynamic limit 
when the $\chi_l^p(t)$'s of two lattice sizes $L$ collapse onto the same line. 
 The NER properties can be inferred from these $\chi_l^p(t)$, 
or we can examine the critical growth of $\chi_l^p(t)$ in the linear region
of a $log(\chi_l^p(t))$ versus $log(t)$ plot. 
This range is called the algebraic range and its upper bound 
is denoted by $\tau$. 
At $T = 1.10J$, $\tau$ appears to increase with $L$ implying that
$\tau \rightarrow \infty$ 
as $L \rightarrow \infty$.
On the other hand, $\chi_l^p(t)$
at $T = 1.25J$ starts saturating for smaller $t$. 
The different behaviors of $\tau$ between these two temperatures 
become more conspicuous in the spin overlap function $q(t)$ (see Appendix).  
However, this speculation requires confirmation in
 complementary investigations.

%%%%%%%%%%%%%%%%%%%%%%%%%%%%%%%%%%%%%%%%%%%%%%%%%%%%%
\begin{figure}
\vspace{0.4cm}
\includegraphics{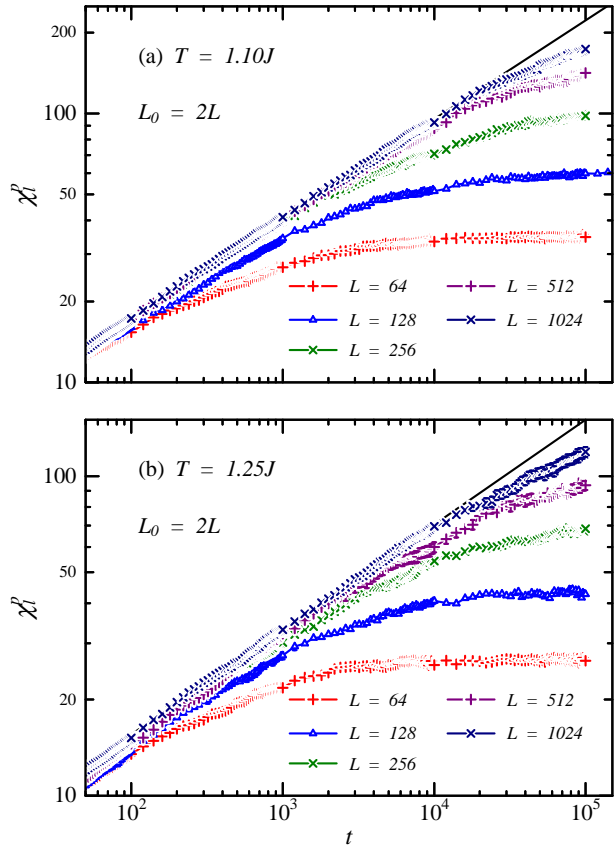}\\
\vspace{-0.4cm}
\caption{ (color online) Growth of $\chi_l^p(t) (\equiv m_l^p(t)L)$ for 
different $L$ at (upper panel) $T = 1.10J$ (slightly lower than $T_{c1}$)
and (lower panel) $T = 1.25J$ (slightly higher than $T_{c1}$).
The straight line in each plot 
is least-squares fitted to the data of $L = 1024$ from 
$50 \leq t \leq 5,000$. 
}
\end{figure}
%%%%%%%%%%%%%%%%%%%%%%%%%%%%%%%%%%%%%%%%%%%%%%%%%%%%%%%%%%%%%%%%%%%%

\subsection{Autocorrelation function in the IC phase}

As shown above, the rapid decay of $m_l^a(t)$ nor $q^a(t)$ does not reveal the 
instability of the IC phase. Here we examine the system dynamics in the IC phase
by the following procedure. 
 First we construct $N (= 64 - 128)$ 
equilibrium spin configurations $\{S_{xy}^{(n)}(0)\}$ $(n = 1 - N)$ at a specified 
temperature $T$, varying the initial spin configurations.
Starting from these equilibrium spin configurations, 
 we conduct MC simulations using the SSF algorithm and obtain the spin configurations 
$\{S_{xy}^{(n)}(t)\}$ at the $t$-th MC sweep. 
The dynamics of the evolving spin configurations are obtained 
from the autocorrelation function $C(T,t)$ defined as  
%%%%%%%%%%%%%%%%%%
\begin{eqnarray}
C(T,t)=\frac{1}{N}\sum_n \frac{1}{L^2}\sum_{x}\sum_{y}S_{x,y}^{(n)}(0)S_{x,y}^{(n)}(t). 
\end{eqnarray}
%%%%%%%%%%%%%%%%
At $t = 0$, $C(T,0) = 1$. 
As $t \rightarrow \infty$, $C(T,t) \rightarrow \langle S\rangle ^2$, where $\langle S\rangle$ 
is the average of $S_{x,y}$ in the equilibrium state.
Then, $C(T,t)$ converges to some positive value at $T < T_c$, 
but exponentially decays at $T > T_c$.  
 At $T = T_c$ or in a critical state, $C(T,t)$ undergoes 
algebraic decay. In other words, $C(T,t)$ plays the same role as the order parameter
 $Q(t)$ in the 
NER simulation.  We now calculate $C(T,t)$ at two $T$ for different $L$.

%%%%%%%%%%%%%%%%%%%%%%%%%%%%%%%%%%%%%%%%%%%%%%%%%%%%%
%\begin{figure}
%\includegraphics{fig13_T110}\\
%\vspace{-0.4cm}
%\caption{ (color online) 
%Autocorrelation function $C(T,t)$ in the ANNNI model with $\kappa = 0.6$ 
%for different $L$ in the equilibrium state at $T = 1.10J$. The straight line 
%is least-squares fitted to the data of $L = 256$ from $50 \leq t \leq 5,000$. 
%}
%\end{figure}
%%%%%%%%%%%%%%%%%%%%%%%%%%%%%%%%%%%%%%%%%%%%%%%%%%%%%%%%%%%%%%%%%%%%

%%%%%%%%%%%%%%%%%%%%%%%%%%%%%%%%%%%%%%%%%%%%%%%%%%%%%
\begin{figure}
\includegraphics{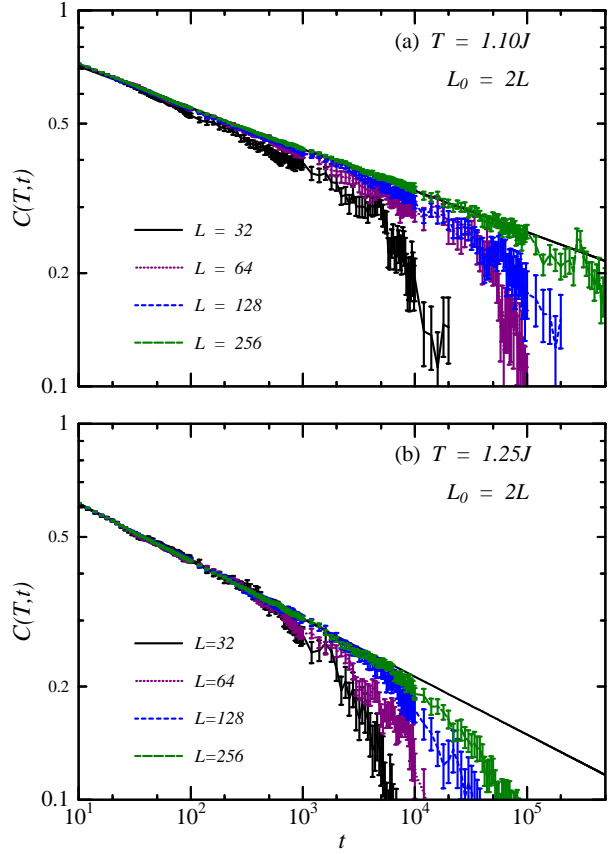}\\
\vspace{-0.4cm}
\caption{ (color online) 
Autocorrelation function $C(T,t)$ in the ANNNI model with $\kappa = 0.6$ 
for different $L$ in the equilibrium state at (upper panel) $T = 1.10J$ and
(lower panel) $T = 1.25J$. The straight line 
is least-squares fitted to the data of $L = 256$ from $50 \leq t \leq 5,000$.
}
\end{figure}
%%%%%%%%%%%%%%%%%%%%%%%%%%%%%%%%%%%%%%%%%%%%%%%%%%%%%%%%%%%%%%%%%%%%

The time courses of $C(T,t)$ at $T = 1.10J$ and $T = 1.25J$ are plotted 
in Figs. 13 (a) and (b), respectively. 
The $L$ dependence of $C(T,t)$ differs between these two temperatures. 
At $T = 1.10J$, the algebraic range extends as $L$ increases, 
while for $T = 1.25J$ it apparently terminates at $t \approx 5000$. 
We make a least-squares fitting to
the function $C(T,t) = At^{-\lambda}$ for the data of $L=256$ from
$50 \leq t \leq 5,000$ and  
the upper bound of the algebraic range $\tau$ is estimated.\cite{tau} 
 Figure 14 plots $\tau$ as a function of $L$ at both temperatures. 
At $T = 1.10J$, $\tau$ appears to extend as $\tau \propto L^z$ with 
$z \approx 2$, while at $T = 1.25J$ it seems to saturate. 
This suggests that $1.10J < T_{c1} < 1.25J$.

%%%%%%%%%%%%%%%%%%%%%%%%%%%%%%%%%%%%%%%%%%%%%%%%%%%%%
%\begin{figure}
%\includegraphics{fig15_time}\\
%\vspace{-0.4cm}
%\caption{ (color online) 
%Algebraic time ranges $\tau$ of the correlation function $C(T,t)$ as functions 
%of linear lattice size $L = L_0/2$ at $T = 1.10J$(black)  and $T = 1.25J$(red). 
%}
%\end{figure}
%%%%%%%%%%%%%%%%%%%%%%%%%%%%%%%%%%%%%%%%%%%%%%%%%%%%%%%%%%%%%%%%%%%%

%%%%%%%%%%%%%%%%%%%%%%%%%%%%%%%%%%%%%%%%%%%%%%%%%%%%%%%%%%%%%%%%%%%%%%%
%%%%%%%%%%%%%%%%%%%%%%%% New_1  use data of fig15_time  %%%%%%%%%%%%%%%
%%%%%%%%%%%%%%%%%%%%%%%%%%%%%%%%%%%%%%%%%%%%%%%%%%%%%%%%%%%%%%%%%%%%%%%

\begin{figure}
\includegraphics{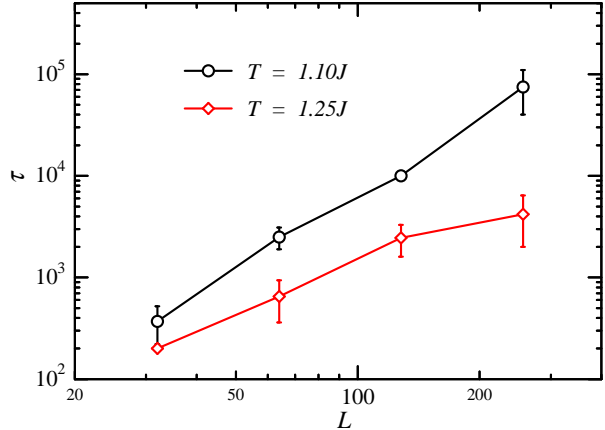}\\
\vspace{-0.4cm}
\caption{ (color online) 
Algebraic time ranges $\tau$ of the correlation function $C(T,t)$ as functions 
of linear lattice size $L = L_0/2$ at $T = 1.10J$(black)  and $T = 1.25J$(red). }
\end{figure}
%%%%%%%%%%%%%%%%%%%%%%%%%%%%%%%%%%%%%%%%%%%%%%%%%%%%%%%%%%%%%%%%%%%%

%%%%%%%%%%%%%%%%%%%%%%%%%%%%%%%%%%%%%%%%%%%%%%%%%%%%%%%%%%%%%%%%%%%%%%%
%%%%%%%%%%%%%%%%%%%%%%%% New_2 use data of Figs. 13 and 24 %%%%%%%%%%%%
%%%%%%%%%%%%%%%%%%%%%%%%%%%%%%%%%%%%%%%%%%%%%%%%%%%%%%%%%%%%%%%%%%%%%%%
%\begin{figure}
%\includegraphics{T_RNG_MA}\\
%\vspace{-0.4cm}
%\caption{ (color online) 
%Algebraic time ranges $\tau$ of the correlation function $C(T,t)$ as functions 
%of linear lattice size $L = L_0/2$ at $T = 1.10J$(black)  and $T = 1.25J$(red). }
%\end{figure}
%%%%%%%%%%%%%%%%%%%%%%%%%%%%%%%%%%%%%%%%%%%%%%%%%%%%%%%%%%%%%%%%%%%%

\subsection{Summary }

Examining the results of recent NER studies on the ANNNI model, 
we find that claims of the absence of the IC phase are not convincing.  

We have reexamined the growth of the IC phase from the PM phase 
by observing the behaviors  
of the chain magnetization $m_l$ and the spin overlap $q$. Both quantities 
of $m_l^p(t)L$ and $q^p(t)L$
appear to algebraically increase with $t$ at $T = 1.10J$. 
We also investigated the autocorrelation function $C(T,t)$ in 
the equilibrium state and found that it algebraically and exponentially
decays over time 
at $T = 1.10J$ and $T = 1.25J$, respectively. In stark contrast to the previous
reports, we conclude that  
the NER method predicts the occurrence of the IC phase below $T_{c1}$ with 
$1.10J < T_{c1} < 1.25J$.

\section{Conclusion}

The spin structure of the 2D ANNNI model is a renewed problem because 
the spin ordering picture in recent large 
scale Monte Carlo (MC) simulations  
depends on the simulation method. 
Specially, the equilibrium simulation predicts a floating 
incommensulate (IC) phase of Kosterlitz-Thouless (KT) type, whereas 
the non-equilibrium relaxation (NER) simulation predicts the absence of this phase. 
In this paper, we examined recently published results of equilibrium and NER 
simulations and 
investigated the spin ordering of the model with frustration ratio $\kappa = 0.6$ 
in both simulation methods. Both methods yielded a 
 KT type 
phase transition between the paramagnetic phase and the IC phase at 
$T_{c1} \approx 1.16J$. 

The present paper focused on the upper phase 
transition at $T_{c1}$. The other phase transition at $T_{c2}$, between the IC phase 
and the $\langle 2 \rangle$ phase, will be investigated in a separate paper.

% If you have acknowledgments, this puts in the proper section head.
\begin{acknowledgements}

We are thankful for the fruitful discussions with Professor S. Fujiki. 
Part of the results in this research was obtained
 using supercomputing resources
at Cyberscience Center, Tohoku University.
% put your acknowledgments here.
\end{acknowledgements}

\appendix
\section{}

%%%%%%%%%%%%%%%%%%%%%%%%%%%%%%%%%%%%%%%%%%%%%%%%%%%%%%%%%%%%%%%%%%%
\begin{figure}
\includegraphics{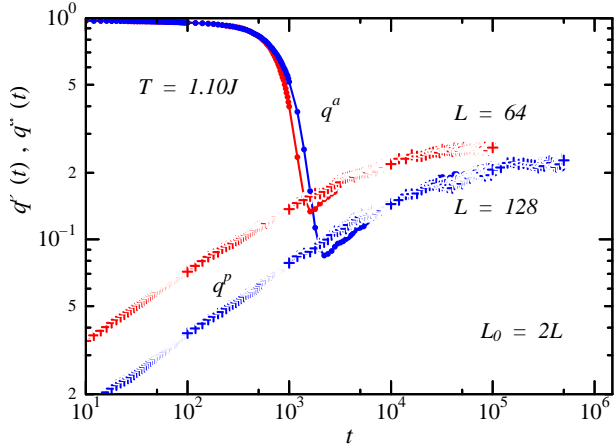}\\
\vspace{-0.4cm}
\caption{(color online) Relaxations of $q^a(t)$ and $q^p(t)$ in the ANNNI model 
with $\kappa = 0.6$ at $T = 1.1J$ (slightly lower than $T_{c1}$). }
\end{figure}
%%%%%%%%%%%%%%%%%%%%%%%%%%%%%%%%%%%%%%%%%%%%%%%%%%%%%%%%%%%%%%%%%%%%

The spin overlap $q^{\mu}(t)$ is calculated as 
%%%%%%%%%%%%%%%%%
\begin{eqnarray}
q^{\mu}(t)=\frac{2}{N(N-1)}\sum_{m\neq n}|\frac{1}{L^2}\sum_{x=1}^{L}
\sum_{y=1}^{L}S_{x,y}^{(m)}(t)S_{x,y}^{(n)}(t)|, \hspace{0.5cm} 
\end{eqnarray}
%%%%%%%%%%%%%%%%%
where $N (= 64 - 256)$ is the number of spin configuration sets 
\{$S_{x,y}^{(n)}(t)\}$ ($n, m = 1 - N$) and $t$ is the MC sweep.   
Figure 15 plots the time courses of
 $q^p(t)$ and $q^a(t)$ with $L_0 = 128$ and 
$L_0 = 256$ at $T = 1.10J$. The temporal behaviors are quite similar to those 
of $m_l^p(t)$ and $m_l^a(t)$ in Fig. 11.

Figure 16 plots the time courses of $q^p(t)L$ for different $L$ at  
$T = 1.10J$(upper panel) and $T = 1.25J$(lower panel). 
At $T = 1.10J$ and $L=1024$, $q^p(t)L$ 
algebraically grows up to $t = 5\times10^4$, 
 whereas at $T = 1.25J$ it starts to saturating at 
$t > 5\times10^3$. 

%%%%%%%%%%%%%%%%%%%%%%%%%%%%%%%%%%%%%%%%%%%%%%%%%%%%%
\begin{figure}
\includegraphics{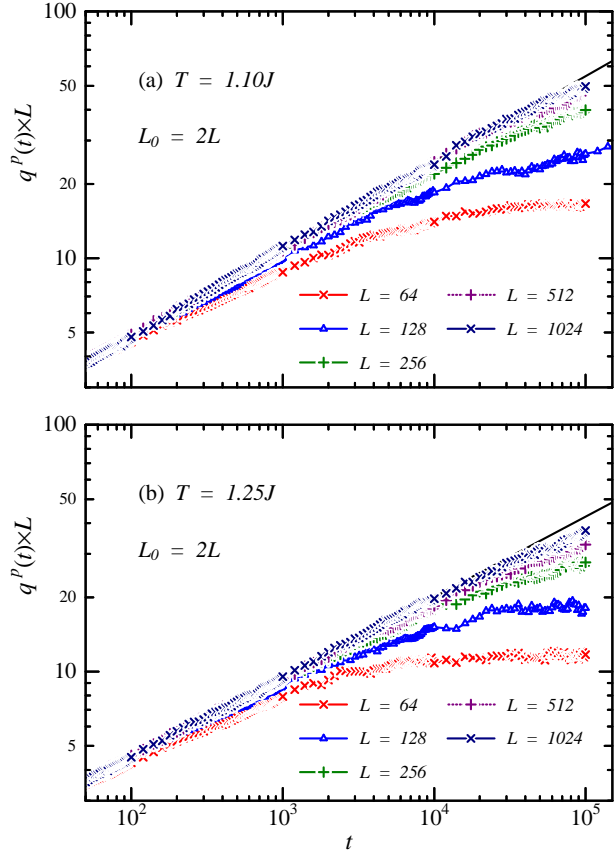}\\
\vspace{-0.4cm}
\caption{(color online) Growth of $q^p(t)$ in the ANNNI model with $\kappa = 0.6$ 
at $T = 1.10J$(upper) and $T = 1.25J$(lower)
 for different $L$. 
The straight line in each plot is least-squared fitted 
to the data of $L = 1024$ from $50 < t < 5,000$.}
\end{figure}
%%%%%%%%%%%%%%%%%%%%%%%%%%%%%%%%%%%%%%%%%%%%%%%%%%%%%%%%%%%%%%%%%%%%

% Create the reference section using BibTeX:
%%%%%%%%%%%%%%%%%%%%%%%\bibliography{basename of .bib file}

\begin{thebibliography}{99}
\bibitem{tau} We roughly estimate $\tau$ as follows. 
We first obtain the fitting function of $At^{-\lambda}$ 
using the least-squares method for the data of $L=256$ from 
$50 \leq t \leq 5,000$, and the relative difference 
$d(t) = |At^{-\lambda} - C(T,t)|/At^{-\lambda}$ is calculated. 
When $t$ is increased from a small $t$, $d(t)$ increases with fluctuation. 
The algebraic ranges $\tau_1$ and $\tau_2$ are defined such that $t's$ 
at which $d(t)$ firstly becomes $d(t) > 0.025$ and $d(t) > 0.05$, respectively.
 We estimate $\tau = (\tau_1 + \tau_2)/2$.
\end{thebibliography}

%

\end{document}